\documentclass{iopconfser}
\usepackage[T1]{fontenc} 
\usepackage{tikz}
\usepackage{braket}
\usepackage{bbm}
\usepackage{mathabx}
\usepackage{mathrsfs}
\usepackage{comment}

\definecolor{indigo(dye)}{rgb}{0.0, 0.25, 0.42}

\newcommand{\beq}{\begin{eqnarray}}
\newcommand{\eeq}{\end{eqnarray}}
\newcommand{\beqn}{\begin{eqnarray}}
\newcommand{\eeqn}{\end{eqnarray}}

\newcommand{\defn}{\mathrel{\mathop:}=} 

\begin{document}
\title{Black Holes, Entanglement and Decoherence}

\author{Gautam Satishchandran$^{1}$}

\affil{$^1$Physics Department, Princeton Gravity Initiative, Princeton, NJ, USA}

\email{gautam.satish@princeton.edu}
\begin{abstract}
It was recently shown that a black hole --- or any Killing horizon --- will decohere any quantum superposition in their vicinity. I review three distinct but equivalent arguments that illustrate how this phenomenon arises: (1) entanglement with ``degrees of freedom'' in the interior (2) the absorption of soft, entangling radiation emitted by the superposition and (3) interactions with the quantum, fluctuating multipole moments of a black hole arising from ultra low frequency Hawking quanta. The relationship between ``soft hair'' and interactions with ``internal degrees of freedom'' is emphasized and some implications for the nature of horizons in a quantum theory of gravity are discussed. 
\end{abstract}
\section{Introduction}
Despite over a hundred years since their discovery, black holes remain as mysterious and enigmatic as ever. Nevertheless, these objects have also provided some of the deepest clues into the quantum nature of the gravitational field. Perhaps the most significant clue has come from the fact that, semiclassically, black holes are not really black! They radiate with a nonvanishing temperature \cite{Hawking:1975vcx} and, therefore, have a finite entropy \cite{Bekenstein:1973ur}. This result has led many to argue that black holes should contain {\em internal degrees of freedom}. The precise origin and counting of these ``microstates'' has been the subject of innumerable investigations and significant debate over the past half century \cite{Gibbons:1976ue,Wald:1993nt,Bombelli:1986rw,Susskind:1994sm,Strominger:1996sh,Ashtekar:1997yu,Maldacena:1996ix,Marolf:2008tx,Chandrasekaran:2022eqq,Kudler-Flam:2023qfl,Climent:2024trz}. In this article we will consider a more modest question concerning the quantum nature of black holes: In what sense does a black hole behave as though it contains such degrees of freedom? What is their impact on physical observables or experiments performed in their exterior? 

In this essay, accompanying a talk given at the GR 24 \& Amaldi 16 conference in July 2025 in Glasgow, I will review some recent results by D. Danielson, R. M. Wald and myself \cite{Danielson:2022tdw,Danielson:2022sga,Danielson:2024yru} which address these questions by considering the entanglement --- and subsequent decoherence --- of a quantum superposition with a black hole. We showed that if a body sources a superposition of gravitational fields in the vicinity of a black hole (or any Killing horizon), then such a body will decohere at a constant rate. Furthermore, the effect of this decoherence is {\em as if} the superposition is becoming entangled with ``internal degrees of freedom'' inside the black hole. The purpose of this article is to explain, in simple terms, the mechanism and origin of this effect. 
In Sections \ref{sec:Bob}, \ref{sec:soft}, and \ref{sec:fluc}, I will present three distinct but equivalent ways of understanding this phenomenon, each highlighting a different aspect of the underlying physics. My goal in focusing on these particular effects is to elucidate what are, in my own view, some universal properties of black holes and horizons in a quantum theory of gravity.

\section{Degrees of Freedom Inside the Black Hole}
\label{sec:Bob}
Classically, black holes have no internal degrees of freedom. Why should a black hole decohere a quantum superposition? The fact that black holes behave in this manner follows straightforwardly from basic principles of quantum mechanics and causality. While the following argument does not appear in the publications mentioned above, it was, in fact, the principal line of reasoning by which we first came to these conclusions.\footnote{This argument, and the conclusions drawn from it, are based on a related gedankenexperiment in flat spacetime, originally proposed in \cite{Mari:2015qva} and analyzed in \cite{Belenchia:2018szb,Danielson:2021egj}. The version presented here was first communicated by me to S. Gralla, and an abridged account appears in \cite{Gralla:2023oya}.} The argument is easiest to present as a gedankenexperiment involving two experimentalists: Alice and Bob.

\par   Alice controls a quantum superposition which she attempts to keeps coherent.  As anyone who's tried to make a quantum superposition knows very well, maintaining coherence is an extremely difficult task! Any degrees of freedom present in her lab will tend to {\em decohere} her superposition. This occurs because the degrees of freedom of the ``environment'' responds differently to each branch of the superposition and will thereby become entangled with it. For example, this could occur due to any interactions with the gas molecules in her laboratory or stray electromagnetic fields from her apparatus. To simply succeed in her experiment she must, at the very least, cool down her lab and electromagnetically shield her apparatus. We will assume that Alice is a first-rate experimentalist and has minimized any decoherence from ordinary environmental effects in her lab. However, Alice cannot perfectly shield the long-range gravitational field of her superposition. If her superposition gravitates differently on one branch as compared to the other then, in principle, any superposition of gravitational fields can be ``measured'' by the environment --- or an external observer --- and will give rise to some decoherence. 

Alice has an adversary, Bob, who wishes to destroy the coherence of her superposition. Despite Alice's best efforts, Bob can achieve this by simply ``measuring'' the superposed gravitational field. Bob could, for example, release a test mass which will accelerate differently depending on the gravitational field of each branch of the superposition. In this way, Alice's superposition will be entangled with the position of his test mass. If his particle becomes sufficiently displaced, then he would have ``measured'' the gravitational field and her superposition will be totally decohered. While Bob can certainly succeed in destroying her coherence, he also doesn't wish to get caught. To ensure that he cannot be convicted for his crimes, Bob decides that he can hide from Alice by measuring her superposition from within a black hole. Of course, the inside of a black hole is by no means an ``optimal'' place to perform an experiment. Nevertheless, the gravitational field of Alice's superposition has no problem penetrating into the black hole interior. Just as before, Bob --- in addition to any number of assistants --- should be able to perform a measurement of the superposed gravitational field. Bob's apparatus must become entangled with Alice's superposition which, by complementarity, must be decohered a commensurate amount. 

However, Alice {\em cannot} know that Bob decided to make a measurement! Despite the fact that they have become entangled, Bob is inside a black hole and his actions cannot have any influence on the outcome of Alice's experiment. To avoid any contradictions with causality and complementarity we conclude that, irrespective of the actions of Bob or his assistants, the black hole itself must decohere Alice's superposition. Furthermore, this decoherence must be {\em as if} the black hole is filled with internal degrees of freedom (i.e., Bob(s)) which entangle with Alice's superposition.

The key feature in the above argument is the existence of an event horizon associated to the worldline of Alice's lab. If Alice's lab is inertial in flat spacetime then there are no event horizons and she can avoid decoherence by performing her experiment sufficiently adiabatically \cite{Belenchia:2018szb,Danielson:2021egj}. However, if her vision of the spacetime is fundamentally limited by an event horizon then, no matter how adiabatically she performs her experiment, her superposition must be decohered at least as much as any degrees of freedom on the other side of the horizon could become entangled with it. In this sense the horizon, itself, ``measures'' a quantum superposition via its superposed gravitational field.



\section{Soft Radiation Absorbed by the Black Hole}
\label{sec:soft}
While considering the implications of possible ``degrees of freedom'' in the black hole interior were very useful in surmising the existence of such an effect, it gives little insight into the mechanism. To gain this insight it will be useful to consider a more specific experimental setup where Alice makes a quantum superposition of spatially separated states in the exterior of a Schwarzschild black hole. In this setup Alice controls a massive body such as a ``nanoparticle'' or some other body whose only relevant degree of freedom is it's center of mass. For brevity, we will refer to this body as a ``particle''. We further assume that the particle has a spin degree of freedom. Given this, one can create a quantum superposition of gravitational fields in the following way. Alice passes the particle through a Stern-Gerlach (or other) apparatus resulting in it being in a superposition state 
\begin{equation}
\frac{1}{\sqrt{2}}(\ket{\psi_{1}}+\ket{\psi_{2}})
\end{equation}
where $\ket{\psi_{1}}$ and $\ket{\psi_{2}}$ are the normalized, spatially separated states of Alice's particle after passing through the apparatus. Alice holds this spatial superposition stationary for a time $T$. After this time, she can check the coherence of her particle by passing it through a reversing Stern-Gerlach apparatus and performing an interference experiment. 
A similar gedankenexperiment involving a spatial superposition of gravitational fields was proposed in the 1950's by Feynman \cite{Dewitt2011TheRO,Zeh:2008dz}. Such gedankenexperiments are now the basis of actual proposed tabletop experiments to measure whether a superposed gravitational field can mediate entanglement \cite{Bose:2017nin,Marletto:2017kzi,Carney:2018ofe,Aspelmeyer:2022fgc}. 

What could be the cause of Alice's decoherence? From her perspective there are no ``Bob(s)'' or ``internal degrees of freedom'' to speak of. Her only source of decoherence can arise from the fact that her particle is coupled to the gravitational field and so may emit some radiation during her experiment. 
Alice can absolutely minimize her radiation to infinity by ensuring that her lab and particle are stationary (i.e., following orbits of the timelike Killing field) and the separation and recombination is done adiabatically.  
However, by the arguments of the previous section, her particle must be forced to radiate in the presence of a black hole. Furthermore, the subsequent entanglement and decoherence must {\em grow} with the time $T$ the superposition is kept open. Since the only other place radiation can go is into the black hole, the final state of Alice's particle must be of the form 
\begin{equation}
\frac{1}{\sqrt{2}}(\ket{\psi_{1}}\otimes \ket{\Psi_{1}}_{\mathscr{H}^{+}}+\ket{\psi_{2}}\otimes \ket{\Psi_{2}}_{\mathscr{H}^{+}})
\end{equation}
where, treating Alice's particle as a test body, $\ket{\Psi_{1}}_{\mathscr{H}^{+}}$ and $\ket{\Psi_{2}}_{\mathscr{H}^{+}}$ are the quantum states of the linearized gravitons which fall through the future horizon $\mathscr{H}^{+}$ of the black hole arising from ``path $1$'' or ``path $2$'' respectively. We have neglected the quantum radiation state at infinity since it produces negligible decoherence. In this simple example, we assume the fluctuations in the stress-energy of the particle are negligible and so the quantum states $\ket{\Psi_{1}}$ and $\ket{\Psi_{2}}$ are well approximated as coherent states. The failure of Alice's particle to interfere is captured by the orthogonality of these states 
\begin{equation}
\mathscr{D}_{BH}\defn 1 - |\braket{\Psi_{1}|\Psi_{2}}_{\mathscr{H}^{+}}| = 1 - e^{-\frac{1}{2}\braket{N}}
\end{equation}
where $\braket{N}$ is the number of gravitons in the difference of the radiation states. We refer to these gravitons as {\em entangling gravitons}. If she emits an O$(1)$ number of entangling gravitons then the radiation states are essentially orthogonal and her particle will be totally decohered. 

We first consider the classical radiation emitted due to slowly displacing a stationary particle initially at a distance $D$ from the black hole by an amount $d$. While the full perturbed metric $h_{ab}$ sourced by the particle in the exterior is quite difficult to explicitly compute, the perturbed metric on the horizon is straightforward to obtain. The (once-contracted) linearized Bianchi identity implies that
\begin{equation}
\label{eq:EABErr}
\mathscr{D}^{A}\mathscr{D}^{B}E_{AB} = \partial_{V}^{2}E_{rr}
\end{equation}
where $E_{ab}=C_{aVbV}$ are components of the linearized ``electric'' Weyl tensor on $\mathscr{H}^{+}$, $V$ is the affine parameter along the horizon, capital Latin indices refer to tensors on the $2$-sphere cross sections of the horizon, $\mathscr{D}_{A}$ is the covariant derivative on the cross-sections and capital indices are raised and lowered with the round metric $q_{AB}$. Physically, $E_{rr}$ encodes the the long-range Coulombic field of the particle and is constant on the horizon when the particle is stationary. $E_{AB}$, on the other hand, encodes the horizon radiation which, by eq.~\ref{eq:EABErr} is non-vanishing when Coulombic field is changing. This change affects the perturbed metric\footnote{We assume that we are in a gauge where $h_{aV}=0=q^{AB}h_{AB}$. These gauge conditions ensure that $\mathscr{H}^{+}$ remains the event horizon in the perturbed spacetime \cite{Hollands:2012sf}.} $h_{AB}$ on the horizon since  $E_{AB} = -\frac{1}{2}\partial_{V}^{2}h_{AB}$. In the process of slowly displacing the particle, negligible energy is radiated into the horizon. Nevertheless, as we will now see, the {\em number} of gravitons emitted into the black hole can be arbitrarily large. Radially displacing the particle by a proper distance $d$ implies that the Coulombic field registered on the horizon changes by an amount $\Delta E_{rr} \sim md^{2}/D^{5}$. Integrating eq.~\ref{eq:EABErr} twice with respect to $V$ we see that, if we had left the particle at the displaced position forever, then the perturbed metric suffers a permanent change on the horizon whose magnitude is proportional to $\Delta E_{rr}$. 

This permanent change of the metric on $\mathscr{H}^{+}$ has been referred to as the ``black hole memory effect'' or ``soft hair'' and was first discussed by Hawking, Perry and Strominger\footnote{Reference \cite{Hawking:2016msc} was primarily concerned with memory created by stress-energy falling through the horizon. Here we see that horizon memory can be generated by the motion of bodies in the black hole exterior.} \cite{Hawking:2016msc} (see also \cite{Rahman:2019bmk,Donnay:2018ckb}). This effect is mathematically equivalent to the memory effect at null infinity whereby changes in momentum of a body --- as opposed to changes in position --- between early and late times result in a permanent change in the metric at leading order in $1/r$ in null directions \cite{1974SvA....18...17Z,Satishchandran:2019pyc}.
In the quantum theory, the memory effect gives rise to infrared divergences arising from the fact that such states contain an infinite number of soft gravitons (see, e.g., \cite{Weinberg:1965nx,Ashtekar:1987tt,Strominger:2013jfa,Strominger:2017zoo,Ashtekar:2018lor,Prabhu:2022zcr,Prabhu:2024lmg}). The number of soft quanta radiated to infinity grows only logarithmically in retarded time \cite{Carney:2017jut} so their effects do not meaningfully affect local experiments which do not occur over extremely long timescales.

Analogously, the horizon memory effect implies that a displaced body emits a number of soft gravitons into the horizon which grows logarithmically in the affine time of the horizon. However, since the affine time duration is exponentially related to total proper time $T$ of Alice's experiment, the number of soft horizon quanta grows {\em linearly} in proper time
\begin{equation}
\braket{N} \sim \frac{M^{5}m^{2}d^{4}}{D^{10}}T \quad \quad \quad \quad \quad \quad \quad \quad \quad\textrm{(GR)}
\end{equation}
where $M$ is the mass of the black hole, $m$ is the mass of the body superposed over a distance $d$ and $D$ is the distance from the horizon. 
Restoring fundamental constants, this gives a decoherence timescale of
\begin{equation}
\label{eq:TDGR}
T_{D}^{GR}\sim \frac{\hbar c^{10}D^{10}}{G^{6}M^{5}m^{2}d^{4}} \sim 1~\textrm{s } \bigg(\frac{D}{\textrm{I.S.C.O.}}\bigg)^{10}\cdot \bigg(\frac{\textrm{M}\textsubscript{\(\odot\)}}{M}\bigg)^{5} \cdot \bigg(\frac{1 \textrm{kg}}{m}\bigg)^{2}\cdot \bigg(\frac{1 \textrm{m}}{d}\bigg)^{4}.
\end{equation}
If Alice's lab was at the inner most stable circular orbit (i.e., $D\sim 6 M$) of a solar mass black hole and she succeeded in superposing a kilogram over a meter, then her superposition would decohere in just 1 second! Of course, in reality, such effects would be totally dominated by decoherence from interactions with the accretion disk and other astrophysical phenomena that we have ignored in this analysis. Furthermore, presently contemplated experiments to create macroscopic quantum superpositions have, at most, considered only nanogram scale masses superposed micrometers \cite{Carney:2018ofe}. Even if one succeeded in creating larger macroscopic superpositions on Earth, the strong dependence on the distance $D$ in eq.~\ref{eq:TDGR} implies that the decoherence from even closest black hole will take longer than the age of the universe unless one considers macroscopic superpositions of astronomical scales. 

Nevertheless, eq.~\ref{eq:TDGR} represents a {\em fundamental} bound on the coherence of any quantum superposition. While the gravitational effect is universal, there are analogous effects for a superposition of any body coupled to a long-range, massive or massless field. In the electromagnetic case, the decoherence timescale of a charge $q$ seperated by a distance $d$ due to the presence of a black hole is
\begin{equation}
 T_{D}^{EM}\sim \frac{\epsilon_{0}\hbar c^{6}D^{6}}{G^{3}M^{3}q^{2}d^{2}} \sim 5~\textrm{min } \bigg(\frac{D}{\textrm{I.S.C.O.}}\bigg)^{6}\cdot \bigg(\frac{\textrm{M}\textsubscript{\(\odot\)}}{M}\bigg)^{3} \cdot \bigg(\frac{\textrm{e}}{q}\bigg)^{2}\cdot \bigg(\frac{\textrm{m}}{d}\bigg)^{2}
\end{equation}
where if Alice had an electron superposed over a meter and her lab was at $D\sim 6 M$ then her coherence would be totally degraded due to the presence of the black hole in about $5$ minutes. Of course, the dependence on distance again implies that current Earth-based interference experiments need not worry about black holes as an important source of ``background''. 

As the arguments of the previous section suggest, 
the entanglement and decoherence by black holes can be generalized to arbitrary Killing horizons \cite{Danielson:2022sga}. 
While our original analysis of a Schwarzschild black hole obtained only order of magnitude estimates for the decoherence rate, Gralla and Wei obtained exact decoherence rates for a spatial superposition with scalar or electromagnetic charge in the exterior of a Kerr black hole \cite{Gralla:2023oya}. Surprisingly they found that, in the near extremal limit, the decoherence rate vanishes in the electromagnetic case! How could this be consistent with the arguments of the previous subsection? It was pointed out by these authors that this is a consequence of a ``black hole Meissner effect'' \cite{Bicak_1985} that maximally spinning black holes screen out any external electromagnetic fields. Thus, for any ``Bob(s)'' inside the black hole, there is no Coulomb field for him to measure, and indeed Alice's particle is not decohered. It would be interesting to incorporate the ``Schwarzian'' corrections to the geometry \cite{Iliesiu:2020qvm} which can affect the absorption cross-section of near extremal black hole at low-frequencies \cite{Emparan:2025sao,Emparan:2025qqf}  and thus affect the decoherence rate \cite{Li:2025vcm,Biggs}. 


Finally, we remark on the relationship between the above arguments and the decoherence inflicted by any ``internal degrees of freedom'' inside the black hole. It can be shown that the amount of ``which path'' information that  can be gained by measuring the superposed, stationary, Coulombic field in the interior is equivalent to the amount of ``which path'' information lost by Alice if she performs her experiment ``optimally'' \cite{Danielson:2021egj,Danielson:2025iud,IDT}. In this sense, these two interpretations are, in fact, equivalent. 
This ``which path'' information carries negligible energy and, indeed, it has recently been shown that this protocol can be used to ``teleport'' information across the horizon with negligible energy cost \cite{Kudler-Flam:2025yur}. 

\section{The Fluctuating Multipole Moments of a Black Hole}
\label{sec:fluc}
 For any physical body with internal degrees of freedom, the entanglement and decoherence of a quantum superposition can be described entirely in terms of local interactions mediated by the long-range field it produces. In the previous sections, we focused on global features of the spacetime—most notably, the presence of a horizon—as the source of decoherence. Here, we provide a fully local account of the same effect, framed entirely in terms of interactions within Alice’s laboratory.

If $T_{1,ab}$ and $T_{2,ab}$ are the classical (expected) stress energies of the components of Alice's superposition on path $1$ and $2$ respectively, then it was shown in \cite{Danielson:2024yru} that the expected number of entangling gravitons emitted by the superposition is equivalently given by the following formula
\begin{equation}
\label{eq:N}
\braket{N} = \braket{\Omega_{U}|h^{in}_{ab}(T_{1}^{ab}-T_{2}^{ab})^{2}|\Omega_{U}}
\end{equation}
where $h^{in}_{ab}$ is the ``unperturbed'', free field operator of the linearized gravitational field which, in eq.~\ref{eq:N}, is smeared in spacetime with the difference of the stress energy of the components of Alice's particle\footnote{In principle, the stress-energy also takes into account the tiny, correlated motion of Alice's lab to keep the center of mass fixed. We assume that Alice's lab is much more massive than her particle so this effect is negligible.} and $\Omega_{U}$ is the ``Unruh vacuum'' --- the relevant vacuum state for a black hole formed from gravitational collapse. From this point of view, the decoherence simply arises from interactions with the {\em vacuum fluctuations} present in Alice's lab. One can obtain a similar formula for the number of emitted entangling photons in terms of interactions of a charged superposition with local electromagnetic vacuum fluctuations \cite{Danielson:2024yru,Wilson-Gerow:2024ljx}. 



The dominant contribution to eq.~\ref{eq:N} arises from the low-frequency structure of the quantum state around a black hole formed from collapse. Unlike an ordinary thermal bath in Minkowski space \cite{Danielson:2022sga,Wilson-Gerow:2024ljx}, the black hole contains an enormous reservoir of ``soft'' Hawking quanta. While these quanta have negligible energy, they partially penetrate into Alice’s lab before being scattered back into the black hole by the angular momentum barrier. Their density of states diverges at low frequencies, and their coherent influence gives rise to fluctuations that {\em precisely mimic that of a material body}.
In particular, the black hole behaves as if it possesses fluctuating multipole moments. The dominant gravitational contribution is a fluctuating mass quadrupole, with constant power spectrum at low frequencies \cite{Danielson:2024yru}
\begin{equation}
\label{eq:quadrupole}
\Delta |Q_{U}|(\omega)\sim \frac{\sqrt{\hbar}G^{2}M^{5/2}}{c^{5}}\sim 10^{-1}\frac{\textrm{g}\cdot \textrm{m}^{2}}{\sqrt{\textrm{Hz}}}\bigg(\frac{M}{M_\odot}\bigg)^{5/2}.
\end{equation}
Similarly, in the electromagnetic case, the black hole acts as though it has a randomly fluctuating electric dipole moment $\vec{P}_{U}$ with constant power spectrum \cite{Danielson:2024yru}
\begin{equation}
\label{eq:dipole}
\Delta |\vec{P}_{U}|(\omega)\sim \frac{\sqrt{\epsilon_{0}\hbar}G^{3/2}M^{3/2}}{c^{3}}\sim 10\frac{\textrm{e}\cdot \textrm{m}}{\sqrt{\textrm{Hz}}}\bigg(\frac{M}{M_\odot}\bigg)^{3/2}.
\end{equation}
Classically, black holes have no hair \cite{Israel:1967wq,Israel:1967za,Carter:1971zc}. Quantum mechanically, however, they acquire an infinite set of randomly fluctuating gravitational and electromagnetic multipole moments. The presence of such zero frequency, fluctuating multipoles in Alice's lab {\em stimulates} the emission of soft photons and gravitons into the black hole. In the absence of such modes --- such as in the stationary vacuum of a static star \cite{Danielson:2024yru} or in a thermal bath --- no such decoherence occurs.

This interpretation permits a direct comparison between the quantum effects of black holes and those of ordinary material systems. A suitably engineered body can reproduce the decoherence induced by a black hole, provided that its low-frequency mass quadrupole and electric dipole fluctuations match those of the black hole itself. This possibility was recently examined by Biggs and Maldacena \cite{Biggs:2024dgp}. For a body to absorb and emit low-energy radiation with the same efficiency as a black hole of comparable size and temperature, it must exhibit extremely high resistivity or viscosity. While these conditions can be met in the electromagnetic case, the gravitational case appears to require unphysical properties of matter \cite{Biggs:2024dgp}.


\section{Epilogue}
The gravitational field, itself, can ``measure'' and decohere a quantum superposition. 
We have shown that any horizon behaves as though it contains ``internal degrees of freedom'' which can become entangled with any quantum mechanical bodies in its vicinity. In the case of black holes, this observation appears to be consistent with explicit ``microscopic'' models of quantum mechanical black holes in string theory \cite{Das:1996wn,Gubser:1997se,Biggs:2024dgp}. While such microscopic models are not presently available for general spacetimes with horizons (e.g., cosmological spacetimes), the arguments presented in this article may shed some light on the microscopic description of these spacetimes and their quantum mechanical dual (if it exists). Additionally, in my own view, a surprising outcome of these analyses is the direct connection between the ``horizon memory effect'' and the entanglement with ``degrees of freedom'' in the interior. While memory and infrared divergences on the horizon have  long been known \cite{Hawking:2016msc,Rahman:2019bmk,Donnay:2018ckb}, the  connections between ``soft hair'' on the horizon and  interactions with ``internal degrees of freedom'' appears worthy of further investigation. 
\par To conclude, while I have attempted to present a coherent and self-contained account of the origin and mechanism of the effect first proposed in \cite{Danielson:2022tdw}, I would caution the reader that this is certainly far from the final word on the subject. As I hope I have made clear, these investigations merely scratch the surface of what remains to be understood about black holes and horizons in a full theory of quantum gravity.




\section*{Acknowledgement}
Though the present reflections are in my own words, --- and thus any errors or unfortunate formulations are mine alone! --- my own understanding of the effects described in this essay has formed from numerous discussions with my collaborators and colleagues. In particular, I have greatly benefited from discussions with M. Aspelmeyer, A. Biggs, D. Carney, S. Carot-Huot, Y. Chen, D. Danielson,  S. Gralla, T. Jacobson, J. Kudler-Flam, J. Maldacena, K. Prabhu, R. M. Wald, J. Wilson-Gerow \& H. Wei. This research was
supported by the Princeton Gravity
Initiative at Princeton University.
\bibliography{iopart-num}
\end{document}